\documentclass{jfm_arxiv}
\usepackage{graphicx}
\usepackage{epstopdf}
\usepackage{natbib}
\usepackage{hyperref}
\usepackage{amsfonts}
\usepackage{bm}
\usepackage{subfigure}
\usepackage{float}

\ifCUPmtlplainloaded \else
  \checkfont{eurm10}
  \iffontfound
    \IfFileExists{upmath.sty}
      {\typeout{^^JFound AMS Euler Roman fonts on the system,
                   using the 'upmath' package.^^J}%
       \usepackage{upmath}}
      {\typeout{^^JFound AMS Euler Roman fonts on the system, but you
                   dont seem to have the}%
       \typeout{'upmath' package installed. JFM.cls can take advantage
                 of these fonts,^^Jif you use 'upmath' package.^^J}%
      }
  \else
  \fi
\fi

\ifCUPmtlplainloaded \else
  \checkfont{msam10}
  \iffontfound
    \IfFileExists{amssymb.sty}
      {\typeout{^^JFound AMS Symbol fonts on the system, using the
                'amssymb' package.^^J}%
       \usepackage{amssymb}%

      }{}
  \fi
\fi

\ifCUPmtlplainloaded \else
  \IfFileExists{amsbsy.sty}
    {\typeout{^^JFound the 'amsbsy' package on the system, using it.^^J}%
     \usepackage{amsbsy}}
    {}
\fi

\newcommand\Rey{\mbox{\textit{Re}}}  

\newsavebox{\astrutbox}
\sbox{\astrutbox}{\rule[-5pt]{0pt}{20pt}}

\newcommand{\Dp}{\ensuremath{D_\mathrm{p}}}

\title{Slipping motion of large neutrally-buoyant particles in turbulence}

\author[M.\ Cisse, H.\ Homann, and J.\ Bec]{Mamadou Cisse, Holger
  Homann, and J{\'e}r{\'e}mie Bec}

\affiliation{Laboratoire Lagrange UMR 7293, Universit\'e de
  Nice-Sophia Antipolis,\\ CNRS, Observatoire de la C\^ote d'Azur,
  Bd.\ de l'Observatoire, 06300 Nice, France.}

\begin{document}
\maketitle
\begin{abstract}
  Direct numerical simulations are used to investigate the individual
  dynamics of large spherical particles suspended in a developed
  homogeneous turbulent flow. A definition of the direction of the
  particle motion relative to the surrounding flow is introduced and
  used to construct the mean fluid velocity profile around the
  particle. This leads to an estimate of the particle slipping
  velocity and its associated Reynolds number. The flow modifications
  due to the particle are then studied. The particle is responsible
  for a shadowing effect that occurs in the wake up to distances of
  the order of its diameter: the particle pacifies turbulent
  fluctuations and reduces the energy dissipation rate compared to its
  average value in the bulk. Dimensional arguments are presented to
  draw an analogy between particle effects on turbulence and wall
  flows. Evidence is obtained on the presence of a logarithmic
  sublayer at distances between the thickness of the viscous boundary
  layer and the particle diameter $\Dp$. Finally, asymptotic arguments
  are used to relate the viscous sublayer quantities to the particle
  size and the properties of the outer turbulence. It is shown in
  particular that the skin-friction Reynolds number behaves as
  $Re_\tau \propto (\Dp/\eta)^{4/3}$.
\end{abstract}


\section{Introduction}
Several natural and industrial phenomena require to model the
transport of finite-size and mass particles suspended in a turbulent
incompressible flow. This is for instance the case for air pollutants,
plankton in the ocean, and industrial mixtures. When the particle size
is much smaller than the smallest active scale of the flow (the
Kolmogorov dissipative scale $\eta$ in turbulence) and when their
Reynolds number defined with their relative velocity to the fluid is
sufficiently small, the surrounding flow can be described by the
linear Stokes
equation~\citep[see][]{gatignol1983faxen,maxey1983equation}.  This
approach leads to model small particles in terms of point-particles
for which an equation of motion can be explicitly written. It has then
been shown that inertia is responsible for particle clustering, a
phenomenon usually referred to as preferential concentration, and for
non-trivial dynamical and statistical properties that can be
characterized in terms of the particle Stokes number and of the flow
properties \citep[see, e.g.,][]{toschi2009lagrangian}.  However much
less is known for particles with sizes comparable or larger than
$\eta$, which can typically have Reynolds numbers larger than
unity. Their dynamics can hardly be modeled because writing an
explicit equation of motion requires fully solving the non-linear
Navier--Stokes equation in the vicinity of the particle. To tackle
this problem, one has to make use either of advanced experimental
particle-tracking techniques or of demanding direct numerical
simulations.

Many recent experimental developments aimed at characterizing the
dynamical properties of finite-size, neutrally buoyant particles (with
the same mass density as the fluid).  Detailed measurements by
\cite{qureshi2007turbulent}, \cite{xu2008motion},
\cite{volk2011dynamics} and \cite{zimmermann2011rotational} of the
translation and angular accelerations suggested that, to a large
extent, finite-size effects can be related to known turbulent
properties calculated at a length scale given by the particle
size. Such approaches are thus implicitly assuming that the presence
of the particle is not altering the fine scaling properties of the
velocity and pressure fields in its vicinity.  While in direct
numerical simulations both the particle motion and the surrounding
fluid flow are by essence always known, such simultaneous measurements
require in experiments an astute setup, as those described by
\cite{khalitov2002simultaneous}, by \cite{bellani2012shape} or by
\cite{klein2012simultaneous}.  In any case, tracking experimentally or
numerically a particle together with the surrounding flow requires a
heavy machinery that complicates the obtention of acute statistics. It
can be for instance particularly laborious to determine joint
distributions of the fluid velocity and the particle acceleration for
an isolated particle in a high-Reynolds turbulent flow. Primarily for
that reason, most of the numerical studies have focused either on a
fixed particle in a developed turbulent flow or on the modulation of
turbulence by many particles \citep[see, e.g.,][for a
review]{balachandar2010turbulent}.

In this paper we make use of a pseudo-spectral solver for the
Navier--Stokes equation, associated to an immersed boundary method to
impose no-slip boundary conditions, in order to study neutrally
buoyant spherical particles which are suspended in a developed
turbulent flow and whose diameters $\Dp$ are within the inertial
range. Our focus is mainly on the local modifications of the flow in
the neighborhood of the particle. A first objective is to understand
the instantaneous direction of the particle slip with respect to the
fluid. We propose a definition that is based on the averaged direction
of the fluid flux in several shells surrounding the spherical
particle. This allows us to compute a mean flow around the moving
particle and to estimate an effective particle Reynolds number. We
then investigate the local modifications of the surrounding turbulent
flow due to the presence of the particle. We show that, while kinetic
energy dissipation is enhanced in the boundary layer, the particle is
calming down turbulent fluctuations in its wake up to distances of the
order of its diameter $\Dp$. Scaling arguments are used to understand
the growth of the turbulent fluctuations as a function of the distance
to the particle.  Similarly to usual wall flows, they show the
presence of a logarithmic law involving a friction velocity and a wall
distance that can be used to collapse the data associated to different
particle sizes. Such arguments, once put together with Kolmogorov
scaling for the outer turbulence, can also be used to show that the
particle friction Reynolds number scales as $\Rey_\tau \propto (\Dp
/\eta)^{4/3}$.

The paper is organized as follows. In \S\ref{sec:slip}, after a short
description of our settings and of the numerical method, we introduce
a definition of the instantaneous direction of the particle slip to
obtain the mean flow around it.  The effects of the particle on the
surrounding turbulent fluctuations are then discussed in
\S\ref{sec:turbulence}.  Finally, concluding remarks are encompassed
in \S\ref{sec:conclusion}.

\section{The slip velocity}
\label{sec:slip}

\begin{table}
\centering
\begin{tabular}{cccccccccc}
$N^3$ &$\delta x$&$\nu$ &$u_\mathrm{rms}$& $\varepsilon$ &
$\eta$ &$\tau_\eta$&$L$&$T_L$& $\Rey_{\lambda}$\\
\hline
 $1024^3$&$6.13\cdot 10^{-3}$&$1.8 \cdot 10^{-4}$&$0.19$
&$4.5\cdot 10^{-3}$&$6.0 \cdot 10^{-3}$& 0.20 &1.6& 8.2 &$160$ \\
\end{tabular}
\caption{\label{table} Parameters of the numerical simulations.
  $N^3$: number of collocation points, $\delta x$:
  grid spacing, $\nu$:  kinematic viscosity, $u_\mathrm{rms}$: root-mean-square velocity,
  $\varepsilon$: mean kinetic energy dissipation rate, $\eta
  =(\nu^3/\varepsilon)^{1/4}$: Kolmogorov dissipation length scale, $\tau_\eta =
  (\nu/\varepsilon)^{1/2}$: Kolmogorov time scale, $L = u_\mathrm{rms}^3/\varepsilon$: integral
  scale, $T_L = L/u_\mathrm{rms}$: large-eddy turnover time,
  $\Rey_\lambda = \sqrt{15\,u_\mathrm{rms}L/ \nu}$: Taylor-microscale
  Reynolds number.}
\end{table}
To address numerically the problem of large-particle dynamics in a
turbulent flow, we make use of a standard pseudo-Fourier-spectral
solver of the Navier--Stokes equations in which the no-slip boundary
condition at the particle surface is imposed by an immersed boundary
technique. The neutrally buoyant particle translational and rotational
dynamics is integrated using Newton's equations. Details and
benchmarks of the method for fixed particles can be found
in~\cite{homann2013effect}.  A similar method has been used in
\cite{homann2010finite} to investigate the dynamics of particles with
sizes of the order of the Kolmogorov scale $\eta$. We report here
results on larger particles. Three independent simulations are
performed with three different particle diameters $\Dp \simeq 17$,
$34$, and $67\eta$. In each case, a large-scale forcing is maintaining
the transporting turbulent flow in a statistical steady state with a
Taylor-microscale Reynold number $\Rey_\lambda\approx160$. The
parameters of the simulations are listed in Tab.~\ref{table}. Three
different runs have been performed, each with a single particle of
size $D_p = 0.1 = 17\,\eta = L/16$, $D_p = 0.2 = 34\,\eta =
L/8$, and $D_p = 0.4 = 67\,\eta = L/4$, respectively. The
turbulent quantities reported in Tab.~\ref{table} vary by less than
2\% between the different runs.

\begin{figure}
  \centering
  \subfigure[$\vec{u}\cdot \vec{e}_x$]{\includegraphics[width=0.495\textwidth]{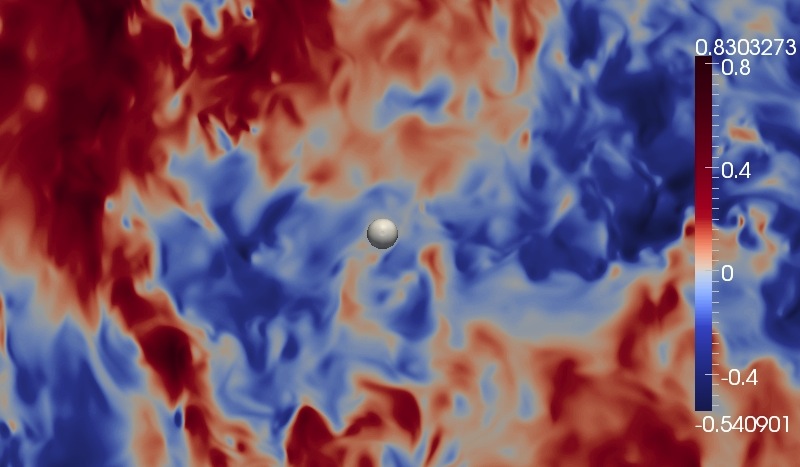}}
  \subfigure[$|\vec{\omega}|$]{\includegraphics[width=0.483\textwidth]{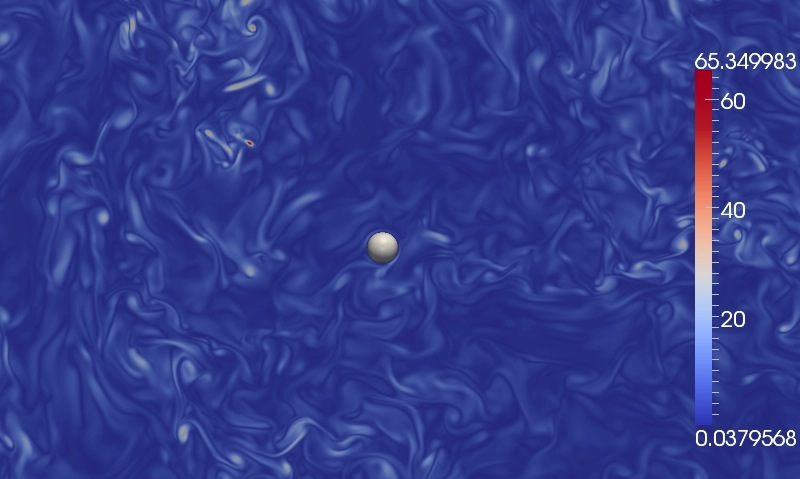}}
  \caption{\label{fig:velocity_cut} (Colour online) Snapshot of one component of the
    velocity (a) and of the vorticity modulus (b) in a thin slice of
    the flow around the particle.}
\end{figure}

Figure~\ref{fig:velocity_cut} shows snapshots (at the same time) of a
component of the fluid velocity (a) and of the modulus of the
vorticity (b) in a plane passing through the center of the particle
for $\Dp = 34\eta$. One clearly sees that in both cases, being
focusing on either large or small-scale fluctuations, the fluid flow
around the particle varies on scales of the order of its size. This
points out one of the key questions in understanding the dynamics of
finite-size particles, that is to define the fluid velocity at the
particle position. This quantity is of particular importance to
evaluate the relative motion (the slip) of the particle with respect
to the carrier flow. All models for particle dynamics consist in
expressing the drag and lift forces exerted by the flow in terms of
this slip velocity. 

\begin{figure}
  \centering
  \includegraphics[width=0.25\textwidth]{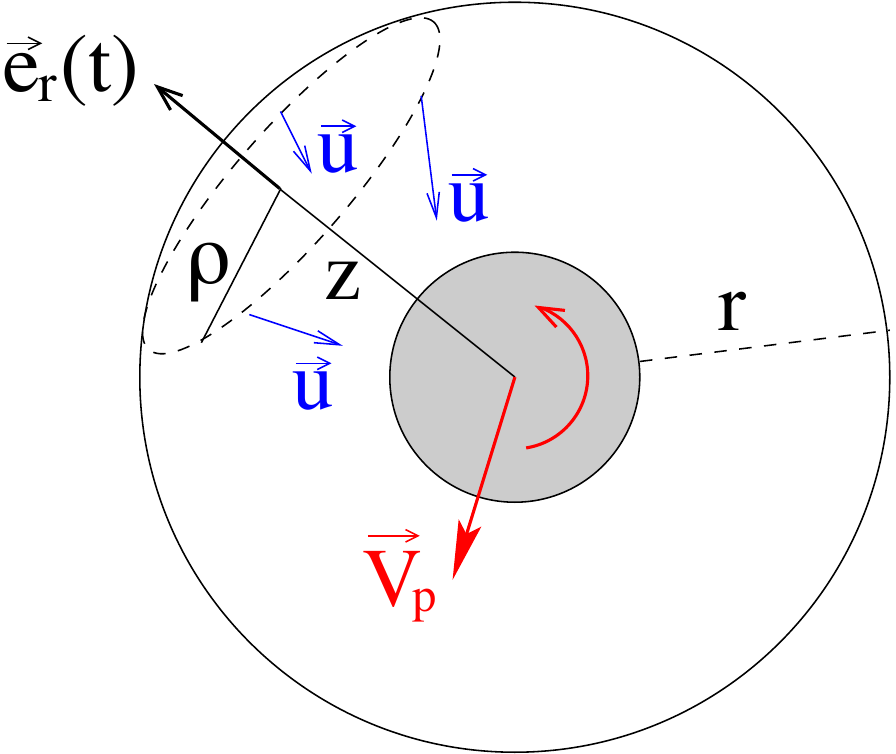}
  \caption{\label{fig:sketch} (Colour online) Sketch of the definition
    of the direction of slip $\vec{e}_r$ on a shell of radius $r$ and of the associated
    instantaneous coordinates $z$ and $\rho$.}
\end{figure}
We aim here at defining an instantaneous direction of the motion of
the particle relative to the fluid. The idea we propose is to evaluate
this direction on different shells surrounding the particle and at
each instant of time. For that we have stored with a sufficiently high
frequency the velocity field in several concentric spheres centered on
the particle. On the shell $\mathcal{S}_r$, which is at a distance $r$
from the particle surface, we define the direction of motion
$\vec{e}_r$ as
\begin{equation}
  \vec{e}_r(t) = \vec{\Phi}_r(t) / |\vec{\Phi}_r(t)|,\quad\mbox{where }
  \vec{\Phi}_r(t) = \int_{\mathcal{S}_r} \left(\vec{u}(\vec{x},t) -
    \vec{V}_\mathrm{p}(t) \right)\cdot\vec{n}\,\,\mathrm{d}\vec{S},
  \label{defer}
\end{equation} 
where $\vec{u}$ and $\vec{V}_\mathrm{p}$ are the fluid and the
particle translational velocity, respectively, $\vec{n}$ is the vector
normal to the shell (see Fig.~\ref{fig:sketch}). In other words, we
perform on each shell an average of the direction weighted by the
fluid mass flux, so that $\vec{e}_r$ points in the direction of the
flux on the shell at distance $r$. This choice is physically motivated
as the fluid enters such a shell upstream and exits in the wake. If
the particle was moving in a laminar flow, the direction $\vec{e}_r$
would be, by symmetry, independent of $r$ and exactly aligned with
this motion. When the particle creates a wake in an unsteady flow, the
direction $\vec{e}_r(t)$ depends on both time and $r$.  Once the
direction $\vec{e}_r$ is defined, one can project on it the velocity
difference $\vec{u}-\vec{V}_\mathrm{p}$ and perform a time average to
construct the mean velocity profile of the flow relative to the
particle
\begin{equation}
  U_\mathrm{rel}(\rho,z) = \left\langle
  \left(\vec{u}(\vec{x},t)-\vec{V}_\mathrm{p}(t) \right)\cdot \vec{e}_r
\right\rangle,
\end{equation}
with $z=(\vec{x}-\vec{X}_\mathrm{p}(t))\cdot\vec{e}_r$ and $\rho =
[\|\vec{x}-\vec{X}_\mathrm{p}(t)\|^2 - z^2]^{1/2}$. The angular
brackets $\langle\cdot\rangle$ designate here the temporal
average. The coordinates $z$ and $\rho$, which are defined at each
instant of time, are in the direction of $\vec{e}_r$ and perpendicular
to it, respectively. By rotational symmetry around the axis defined by
$\vec{e}_r$, the mean profile $U_\mathrm{rel}$ depends on $z$ and
$\rho$ only and not on the angle.

\begin{figure}
  \centering
  \includegraphics[width=0.9\textwidth]{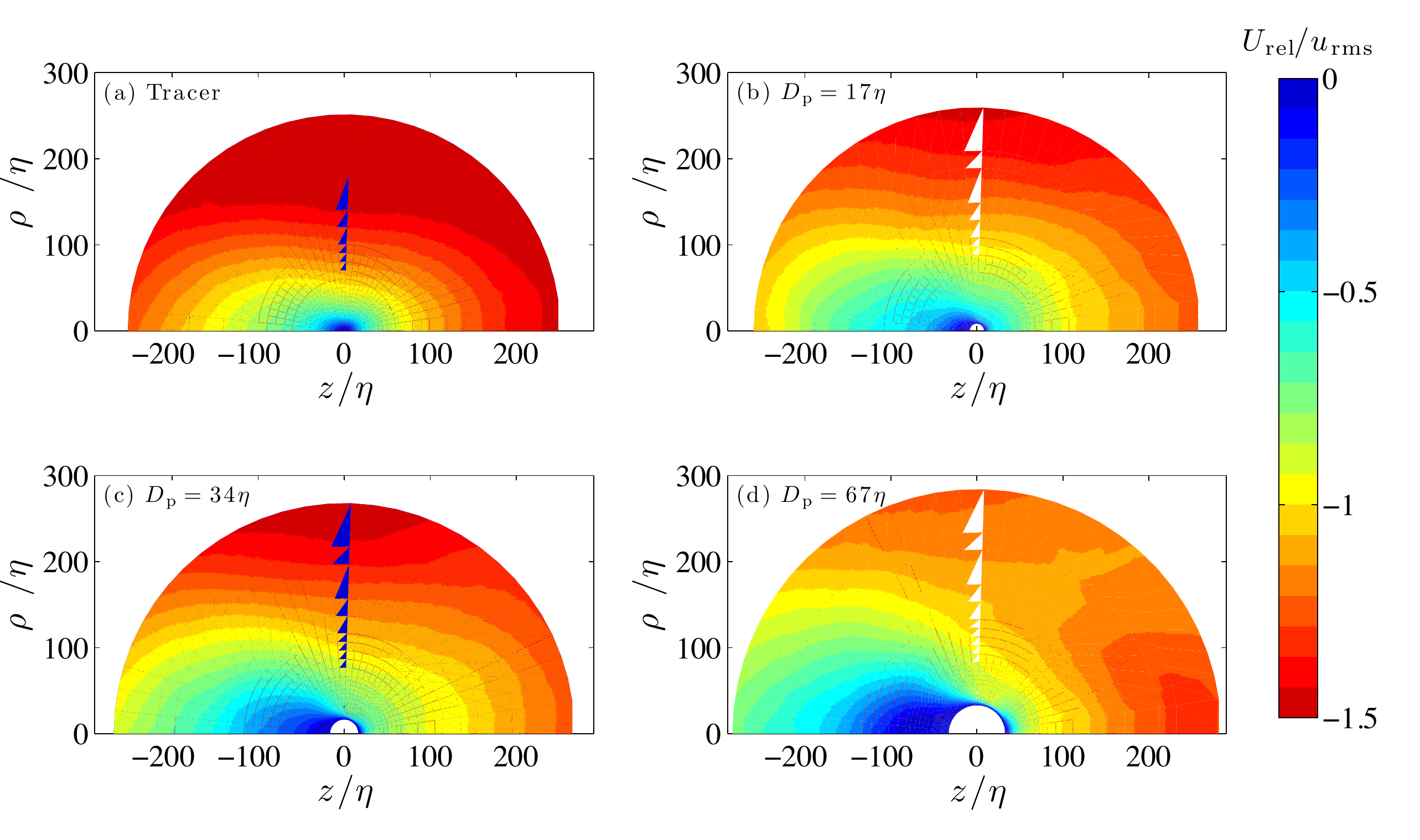}
  \caption{\label{fig:snapshot} (Colour online) Temporal- and
    angle-averaged relative velocity (definition with flux) of the
    fluid projected on the direction of motion for (a) a tracer and
    (b) $\Dp=17\eta$, (c) $\Dp=34\eta$, and (d) $\Dp=67\eta$.}
\end{figure}
Figure~\ref{fig:snapshot} represents the measured average velocity
profile for a tracer and for the three particle sizes. The relative
motion of the fluid with respect to the particle is from $z>0$ to
$z<0$. In all four cases, the upstream and downstream velocities are
clearly asymmetric. Also, when the particle radius increases, one
observes the development in the wake of a region where the flow is
calmed down. However a large part of the information contained in the
mean relative velocity $U_\mathrm{rel}$ is purely due to
kinematics. This is clear when interested in the case of the tracer:
the surrounding flow is trivially not affected by its presence but the
conditioning in terms of the instantaneous flux direction $\vec{e}_r$
prevents the average relative velocity profile from vanishing and
singles out the growth of turbulent velocity increments $\sim
r^{1/3}$. The observed asymmetry relates to the fact that the negative
longitudinal velocity differences (on the right) are more likely to be
larger than the positive ones (on the left); this can be interpreted
as a consequence of the 4/5 law and of the resulting skewness of
velocity differences in turbulence.

\begin{figure}
  \centering
  \subfigure[]{\includegraphics[height=0.32\textwidth]{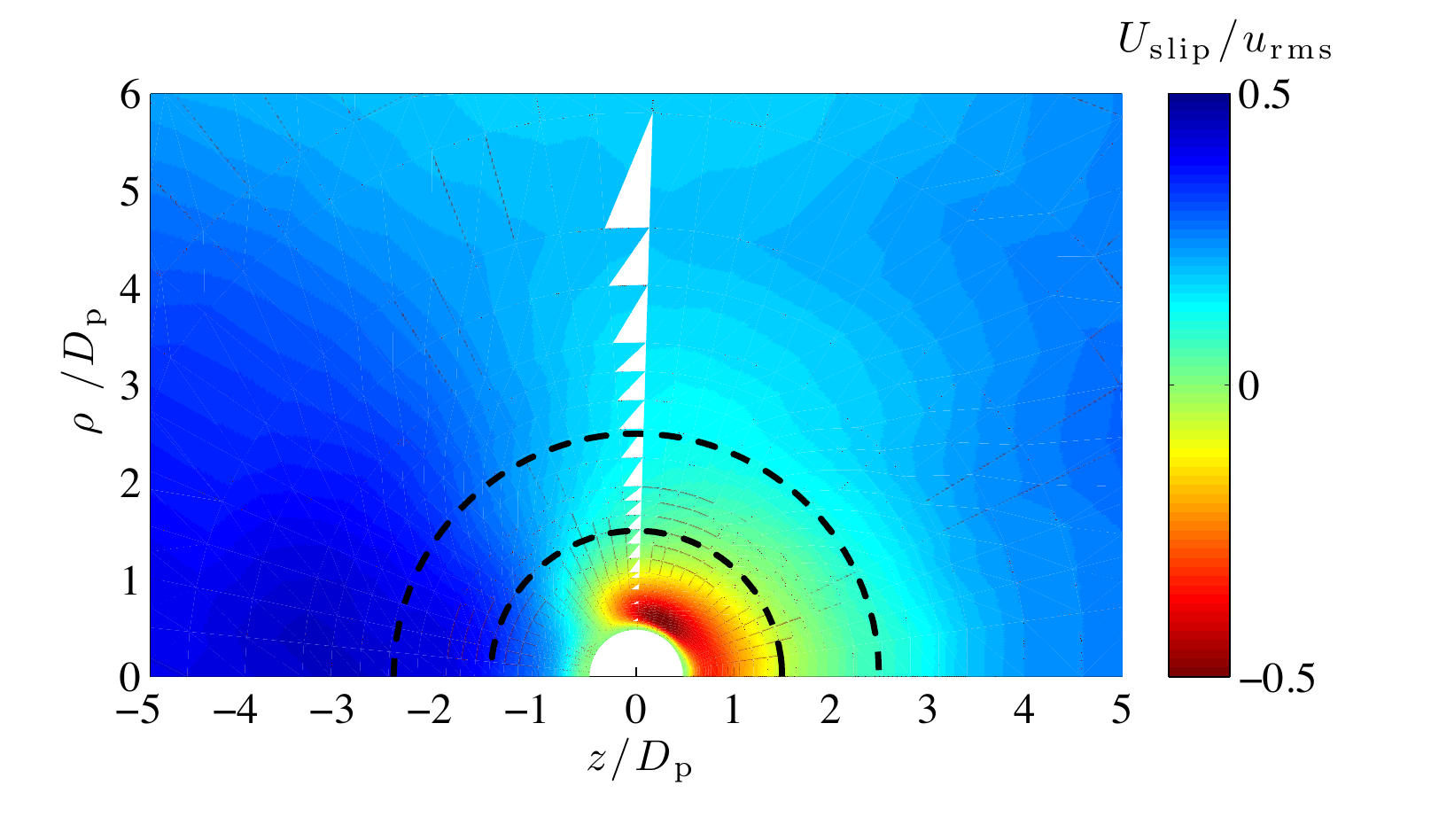}}
  \subfigure[]{\includegraphics[height=0.32\textwidth]{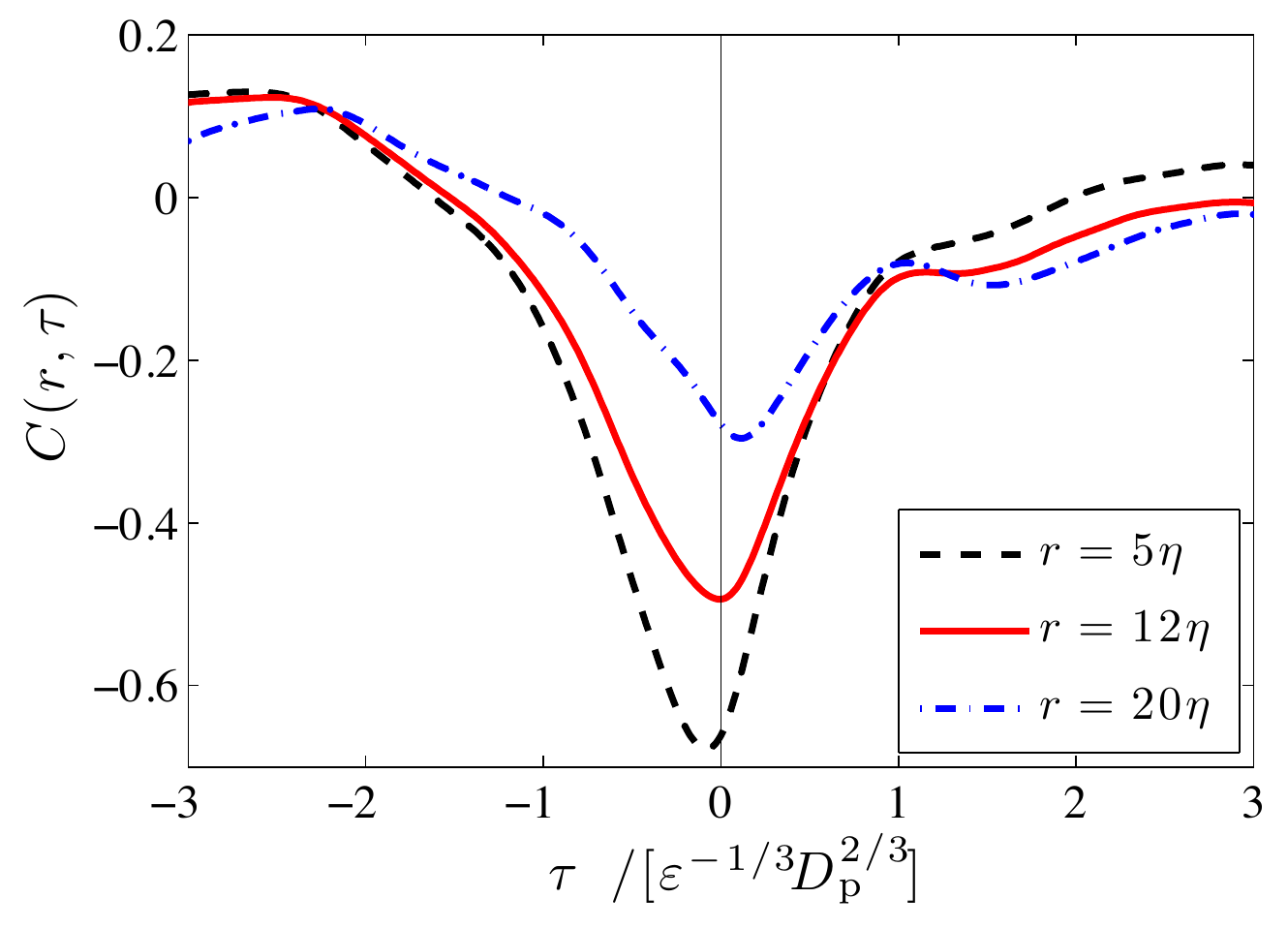}} 
  \caption{\label{fig:CarteVelocityDifference} (Colour online) (a)
    slip velocity $U_\mathrm{slip}$ for $\Dp = 34\eta$ defined as the
    difference between the mean relative velocity profile
    $U_\mathrm{rel}$ around the particle and that around a tracer; the
    two black dashed circles represent distance equal to $\Dp$ and
    $2\Dp$ from the particle surface. (b) Time correlation $C(r,\tau)=
    \langle \vec{F}(t+\tau) \cdot \vec{e}_r(t) \rangle / \langle
    |\vec{F}|^2 \rangle^{1/2}$ between the force at time $t+\tau$ and
    the direction of motion at time $t$ for three values of the shell
    distance $r$ and for $\Dp=34\eta$.}
\end{figure}
Actually, the details of the particle slip and of the flow
modifications due to its presence can be obtained by comparing the
average profiles with a particle to that without it. One observes in
Fig.~\ref{fig:snapshot} that the most noticeable differences occur on
scales of the order of the particle diameter. This is even clearer in
Fig.~\ref{fig:CarteVelocityDifference}(a), which represents for $\Dp =
34\eta$ the ``slip velocity'' profile defined as
\begin{equation}
  U_\mathrm{slip}(\rho,z) = U_\mathrm{rel}(\rho,z) -
  U_\mathrm{rel}^\mathrm{tracer}(\rho,z).
\end{equation}
This quantity is the difference between two velocity differences.  The
four terms it contains can in principle be grouped in two
contributions: the difference between the fluid velocity with and
without the particle, which accounts for the flow modifications due to
its presence, and minus the difference between the particle velocity
and the velocity of a tracer that would be at the particle
location. The first term comprises all the space dependency. One
observes from Fig.~\ref{fig:CarteVelocityDifference}(a) that the flow
modifications due to the particle are up to distances of the order of
its diameter and vanish far from the particle.  At sufficiently large
distances, $U_\mathrm{slip}$ attains a positive constant coming from
the second contribution. This limit, that we denote
$U_\mathrm{slip}^\infty$ can be used to define a difference between
the particle velocity and that of the fluid at the particle location,
that is a typical slip velocity. We can make use of
$U_\mathrm{slip}^\infty$ in order to define a particle Reynolds number
$Re_p=U_\mathrm{slip}^\infty\,\Dp/\nu$. We obtain:\\
\strut\quad \begin{tabular}{ll}
- for $\Dp=17\eta$: & $U_\mathrm{slip}^\infty\approx0.049$ and $Re_p\approx27$, \\
- for $\Dp=34\eta$: & $U_\mathrm{slip}^\infty\approx0.052$ and $Re_p\approx58$, \\
- for $\Dp=67\eta$: & $U_\mathrm{slip}^\infty\approx0.057$ and $Re_p\approx126$.
\end{tabular}\\
The typical turbulent fluid velocity fluctuation $\propto(\varepsilon
\Dp)^{1/3}$ at a separation $\Dp$ is
from 1.5 to 2 times larger than these values of the slip velocity.
Note that the definition of a slip velocity that we are using
here is based on the instantaneous direction of motion $\vec{e}_r$
introduced earlier. It thus differs from the slip definitions based on
statistical arguments, such as that used by \cite{bellani2012slip}.

One step further to assess the validity of the proposed definition of
the relative motion direction $\vec{e}_r$ consists in looking at its
alignement with the force $\vec{F}$ exerted by the fluid onto the
particle. For that we define the correlation $C(r,\tau)= \langle
\vec{F}(t+\tau) \cdot \vec{e}_r(t) \rangle / \langle |\vec{F}|^2
\rangle^{1/2}$ between the force at time $t+\tau$ and the direction of
motion of the shell at distance $r$ and at time $t$. As seen from
Fig.~\ref{fig:CarteVelocityDifference}(b) the two vectors are
anti-correlated. The anti-correlation is maximal very close to the
particle surface where the flow is trivially enslaved to the solid
motion because of the no-slip boundary condition. However, the optimal
lag $\tau$ which maximizes the anti-correlation is there
negative. This means that the force is there imposing the direction
$\vec{e}_r$ of the local motion. The anti-correlation decreases when
$r$ increases and becomes very small when $r\gg \Dp$. At sufficiently
large distances, the minimum of correlation is attained for a positive
value of the time lag $\tau$. It means that sufficiently far from the
particle, the flow direction is in advance on that of the force. There
is thus a specific value of $r$, of the order of $\Dp/2$, for which
the optimal lag vanishes and the direction of the force is almost
synchronized with that of the relative motion. This indicates that the
fluid-particle interactions occur at distances up to the order of the
particle diameter $\Dp$, as already observed by
\cite{naso2010interaction}. Also, we find that a part of the force
exerted by the fluid corresponds to a drag in the direction
$\vec{e}_r$ of the relative motion.

\section{Turbulence statistics in the vicinity of the particle}
\label{sec:turbulence}
\subsection{Imprint on the kinetic energy and the dissipation rate}
\label{sec:imprint}

\begin{figure}
   \centering
   \subfigure[]{\includegraphics[width=0.45\textwidth]{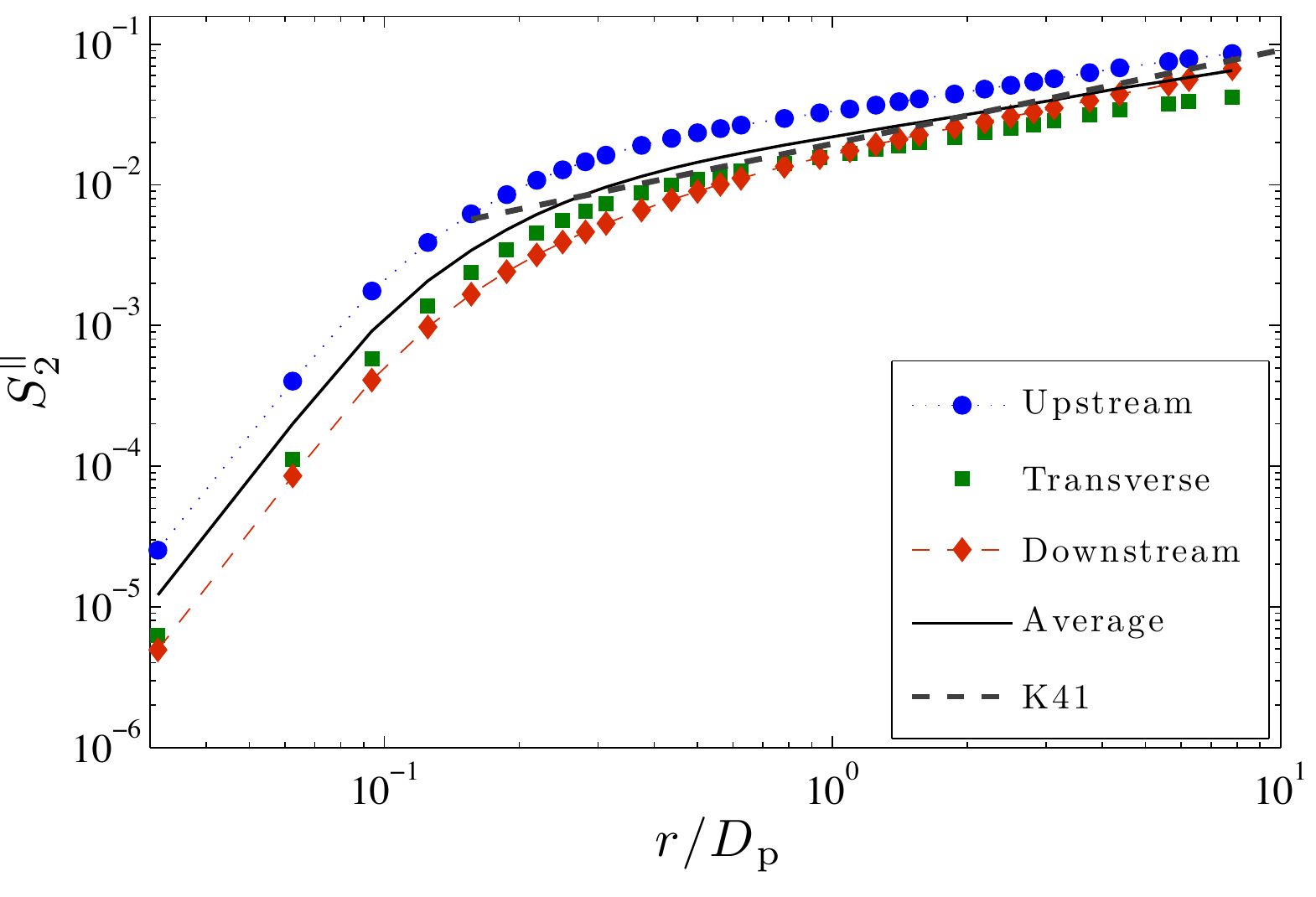}}
    \subfigure[]{\includegraphics[width=0.45\textwidth]{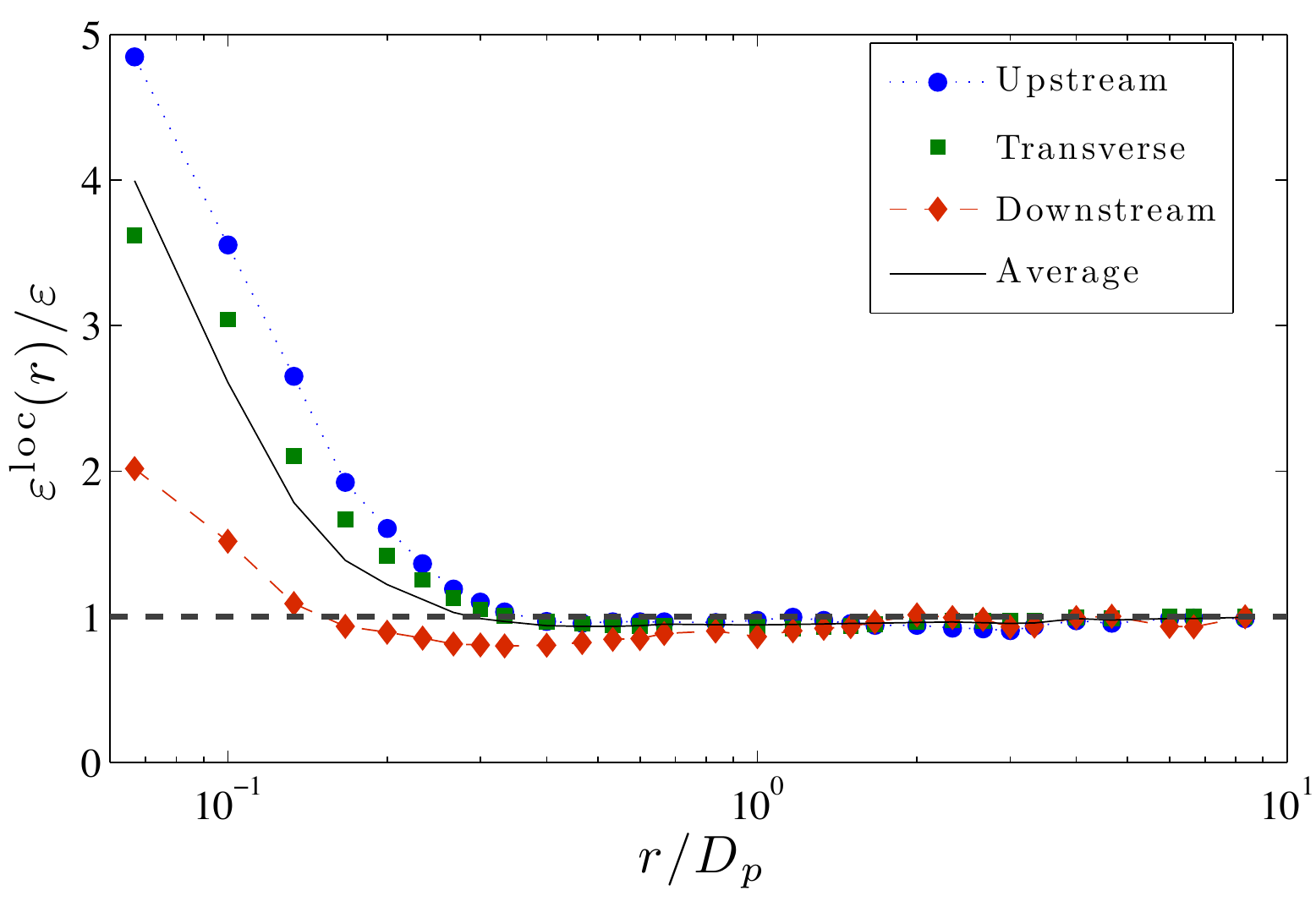}}
    \caption{\label{fig:FslDissipationRate} (Colour online) (a)
      Particle-anchored second order structure function (the K41 line
      stands for $S_2^\parallel = C_2 (\varepsilon r)^{2/3}$ with $C_2
      = 2.1$) and (b) local dissipation rate in different regions
      around the particle as a function of the distance to its surface
      for $D_p=34\eta$. }
\end{figure}

We now turn to the influence of the particle on higher-order
statistics of the fluid velocity
field. Figure~\ref{fig:FslDissipationRate}(a) shows the
``particle-anchored'' second-order longitudinal structure function
defined as
\begin{equation}
  S^\parallel_2(r) = \left\langle \left[ \left(
        \vec{u}(\vec{x},t)-\vec{V}_\mathrm{p}(t) \right)\cdot \vec{n}\right]^2\right\rangle,
\end{equation}
where $\vec{n}$ is the unit vector in the direction of
$\vec{x}-\vec{X}_\mathrm{p}$ and $r = \| \vec{x}-\vec{X}_\mathrm{p}\|
- \Dp/2$ is the distance from $\vec{x}$ to the particle surface. Using
the instantaneous and $r$-dependent definition of the relative motion
direction $\vec{e}_r(t)$ of previous section, the average can be
decomposed in an upstream contribution (for $\vec{x}$ in a cone of
90$^\mathrm{o}$ in the direction of $\vec{e}_r$), a downstream
contribution (in a cone of 90$^\mathrm{o}$ in the direction of
$-\vec{e}_r$) and a transverse contribution (remaining values). One
observes in Fig.~\ref{fig:FslDissipationRate}(a) that the velocity
fluctuations are apparently enhanced upstream the particle. This is
essentially due to the average flow modifications as the flow is
accelerated when approaching the particle and encounters steep
gradients. In principle one would expect a similar behaviour downstream
as the flow is there decelerated. However we observe the reverse
phenomenon as the large-$r$ asymptotics is reached from below. This
indicates that the particle is calming down turbulence in its wake.

This effect is even more visible in the profile of the average kinetic
energy dissipation rate $\varepsilon^\mathrm{loc}(r)$ computed as a
function of the distance $r$ to the particle surface and conditioned
on the upstream, downstream and transversal sectors. As seen from
Fig.~\ref{fig:FslDissipationRate}(b), while energy dissipation is
strongly enhanced in all directions in the vicinity of the particle,
its downstream value is below $90\%$ of its average up to distances of
the order of the particle diameter. This indicates again that
turbulence is more quiet downstream.

The particle is thus creating a shadow in its wake. An explanation
relies on the fact that all turbulent structures of sizes of the order
of $\Dp$ are not anymore present in the downstream flow. In addition,
given the pretty low values of the Reynolds numbers estimated in
previous section, the particle wake is not strong enough to inject a
significant amount of turbulent kinetic energy. This shadowing effect
seems to depend only weakly on the particle size, up to the short
range of values we have investigated and the accuracy of our
simulations.

\subsection{Analogy with wall turbulence}
To analyse more precisely the particle size dependence of the
disturbed flow, we make use of an approach similar to that used in
wall turbulence.  A first important difference is that in the case of
particles the ``bulk flow'' is not a input data, so that our approach
relies on what is happening in the immediate neighborhood of the
boundary. A second difference is that, because of the isotropy of the
particle dynamics, the velocity field averages to zero and there is no
notion of mean velocity profile. For this reason, we make use of the
root-mean-square velocity difference between the flow and the particle
surface in the tangential directions
\begin{equation}
  U(r) = \left\langle \frac{1}{2} \left\| \vec{u}- (\vec{u}\cdot\vec{n})\,\vec{n} -
    \vec{V}_\mathrm{p} - \frac{\Dp}{2}\,\vec{\Omega}_\mathrm{p} \times
    \vec{n}\right\|^2 \right\rangle^{1/2},
\end{equation}
where $\vec{n}$ is the unit vector normal to the spherical particle
and $\vec{\Omega}_\mathrm{p}$ is the particle angular velocity.
$U^2(r)$ is nothing but the second-order ``particle-anchored''
transverse structure function. Our numerical data allows us to measure
a \emph{wall shear stress} as
\begin{equation}
  \tau_\mathrm{w} = \nu \left.\left[ \frac{\mathrm{d}U}{\mathrm{d}r}
      \right ] \right|_{r=0},
\end{equation}
As in the case of wall flows \citep[see, e.g.,][]{pope}, this
quantity, together with the viscosity $\nu$, defines all relevant
quantities of the viscous boundary layer surrounding the particle,
namely the \emph{friction velocity} $u_\tau = \sqrt{\tau_\mathrm{w}}$,
the \emph{viscous lengthscale} $\delta_\nu = \nu/u_\tau$, and the
\emph{friction Reynolds number} $Re_\tau = u_\tau\,\Dp/\nu$. Also, $U$
and $r$ can be written in wall units by introducing $U^+ = U / u_\tau$
and $r^+ = r/\delta_\nu$.

The fluid velocity at a distance $r$ from a particle is completely
determined by $u_\tau$, $\delta_\nu$, $\Dp$, $L$, and
$\Rey_{\lambda}$. Dimensional analysis then suggests to write
\begin{equation}
  \frac{\mathrm{d}U}{\mathrm{d}r}=
  \frac{u_\tau}{r}\,\Psi({r}/{{\delta}_\nu},\,{r}/{{D}_p},\,
  {r}/{L},\Rey_{\lambda}),
\label{equ:anal_dim}
\end{equation}
where we have non-dimensionalized lenghtscales with $r$ and velocities
with $u_\tau$. We next make use of the scale separation
$\delta_\nu\ll\Dp\ll L$ to evidence different layers.
\begin{itemize}
\item $r\ll\delta_\nu$ corresponds to the viscous sublayer where by
  construction $U^+ \simeq r^+$.
\item $\delta_\nu \ll r \ll \Dp$ is the outer layer. We have
  $\mathrm{d}U/\mathrm{d}r \simeq
  (u_\tau/r)\,\Psi_\star(\Rey_\lambda)$, where
  $\Psi_\star(\Rey_\lambda) = \Psi(\infty,0,0,\Rey_\lambda)$. As in
  wall turbulence, this leads to the \emph{log-law}
\begin{equation}
  U^+ = C+\Psi_\star(\Rey_\lambda)\,\ln r^+.
\end{equation}
\item $\Dp\ll r \ll L$ corresponds to distances far from the particle
  where turbulent fluid statistics are recovered. In the limit of very
  large Reynolds numbers we can assume that $r/L\to 0$. At large
  distances from the particle, the behaviour of $U$ should be given by
  the fluid velocity second-order structure function. According to
  Kolmogorov 1941 scaling, we expect $U^2 \simeq (4/3) \,
  C_2\,(\varepsilon r)^{2/3}$. This implies that for $r/\Dp\to\infty$,
  the dimensionless function $\Psi$ has to diverge as a power law
  (with exponent $\alpha$ and a constant $\Psi_\infty$ that depends on
  the outer Reynolds number), so that $\mathrm{d}U/\mathrm{d}r \simeq
  \Psi_\infty (\Rey_{\lambda})\, (u_\tau/r) \, (r/\Dp)^\alpha$.  when
  $r\gg\Dp$. We hence find that $\alpha=1/3$ and $u_\tau \propto
  (\varepsilon \Dp)^{1/3}$.
\end{itemize}

\begin{figure}
   \centering
   \includegraphics[width=0.6\textwidth]{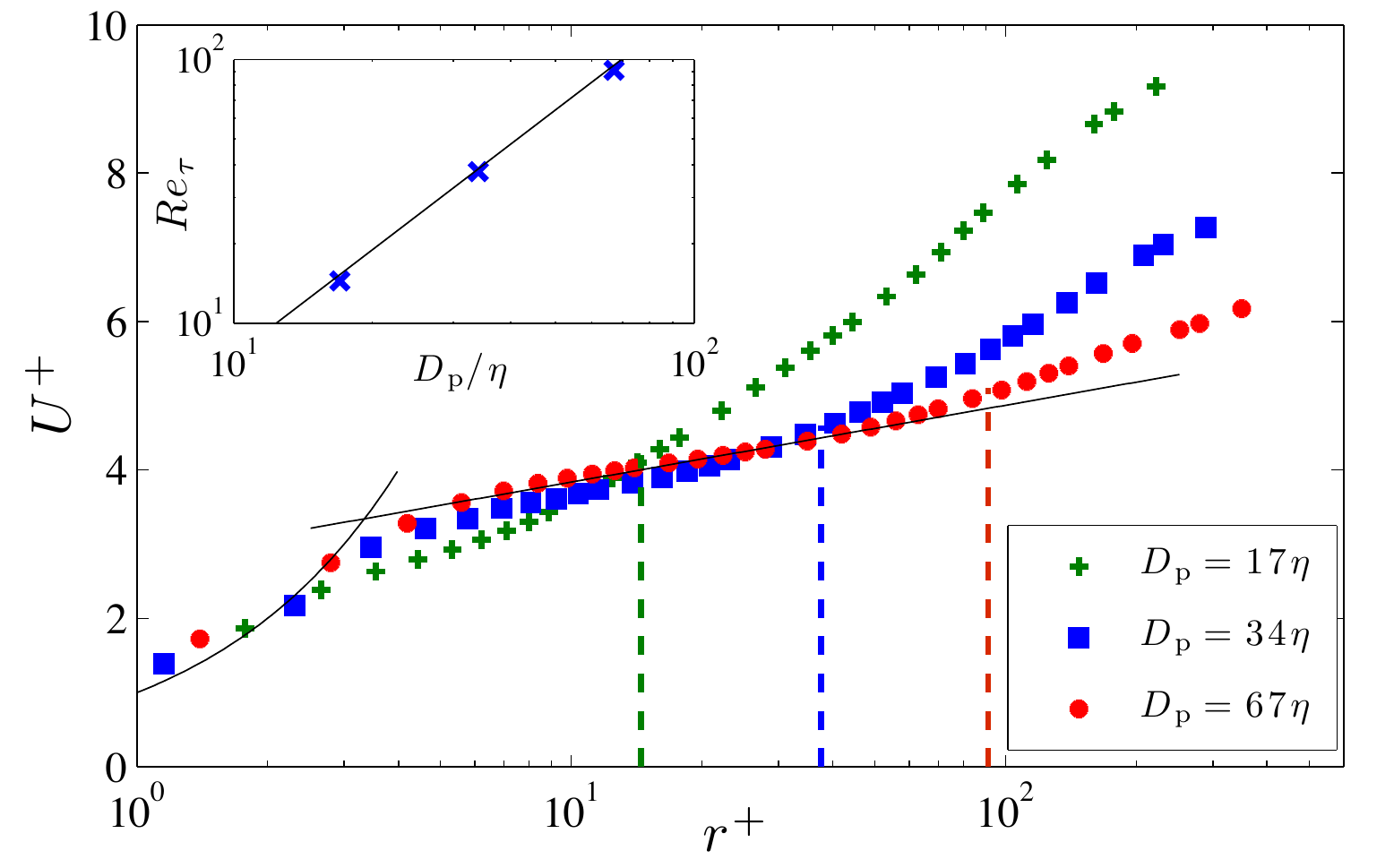}
   \caption{\label{fig:logLaw} (Colour online) Average transverse
     velocity amplitude $U$ as a function of the distance to the
     particle surface in wall units for three different particle
     sizes; the black solid curve on the left refers to the viscous
     sublayer; the black solid line is a fit to the logarithmic law of
     the form $U^+= 2.8+0.45\ln r^+$; the three dashed vertical lines
     represents the values of $\Dp$ in wall units. Inset: Friction
     Reynolds number $Re_\tau$ as a function of the particle diameter
     (crosses); the black line corresponds to $Re_\tau =
     0.35\,(\Dp/\eta)^{4/3}$. }
\end{figure}
These three asymptotics can be observed in Fig.~\ref{fig:logLaw} where
the amplitude $U$ of the tangential velocity difference is represented
as a function of the distance to the particle surface. One clearly
sees a log-law region that becomes wider when $\Dp$ increases.  The
relationship $u_\tau \propto (\varepsilon \Dp)^{1/3}$ that was obtain
by matching the large-$r$ asymptotics to the behaviour of turbulent
structure functions implies that the viscous lengthscale obeys
$\delta_\nu/\eta \propto (\Dp/\eta)^{-1/3}$ and that the friction
Reynolds number depends on the particle size as $Re_\tau \propto
(D_p/\eta)^{4/3}$. This latter behaviour is confirmed numerically as
seen from the inset of Fig.~\ref{fig:logLaw}. Note that these values
differ roughly by a factor 2 from those obtained in \S\ref{sec:slip}
from our estimate of the slip velocity. To conclude this analysis, let
us stress that the simple dimensional arguments developed here and
which seem to be numerically confirmed, show that the important
parameter to specify the fluid flow in the particle neighborhood is
the non-dimensional ratio $\Dp/\eta$.

\section{Concluding remarks}
\label{sec:conclusion}

In this paper, we have investigated the interactions between a large
particle and the turbulent flow surrounding it. We proposed a
definition of an instantaneous direction in which the particle slips
with respect to the fluid using the mass fluxes in different
concentric shells centered on the particle. This definition allowed us
to construct a mean flow around the particle and to define a typical
slip velocity. We next turned to the effect of the particle on the
properties of the surrounding turbulence. We have seen that kinetic
energy dissipation is reduced in the wake, so that particles are
responsible for a kind of shadowing effect on the flow. Finally, we
have presented dimensional arguments analogous to those used for wall
turbulence in order to characterize the velocity fluctuations in the
direction transverse to the particle surface. We have shown the
presence of a log law and related the viscous sublayer properties to
the particle size and the carrier flow turbulence.

A potential application of our work relates to the design of models
for the dynamics of large-size particles suspended in a turbulent
flow. In most practical situations that are encountered in engineering
or atmospheric sciences, the flow is under-resolved (as for instance
in large-eddy simulations) and the dynamics of particles with large
inertial-range sizes still below the cutoff scale is approached by
point particles \citep[see, e.g.,][for more
details]{Balachandar2009scaling}. The force acting on the particle is
then approximated by the standard drag model, possibly including
empirical corrections due to the presence of turbulence in the
particle surrounding. The approach we have proposed here opens new
ways in tackling such issues in terms of slip direction, shell
averages, and log layer. This goes beyond the scope of the present
work as it will require a huge computational investment to study, for
instance, the correlations between the force and the surrounding flow.

Finally, it is important to mention that we have focused here on
isolated large-size particles. Our results do not straightforwardly
extend to the interactions between several of them. However, in very
dilute settings, we expect the modulation of turbulence by a dispersed
phase to be affected by our findings. As seen in \S\ref{sec:imprint},
the change in energy dissipation is two-fold: on the one hand, it is
increased in the immediate vicinity of the particle and, on the other
hand, it is weakened in its wake. This non-uniform effect can have
non-trivial consequences on the coupling between the flow and the
particles, to which possible collective effects can add up. One can
for instance imagine that particles gather even if they are neutrally
buoyant and unaffected by preferential concentration, the main
mechanism being a collective shadowing of turbulent fluctuations that
prevent eddies from separating them.

\medskip

We thank M.\ Gibert for useful discussions. This work was performed
using HPC resources from GENCI-IDRIS (Grant 026174) and from FZ
J\"ulich (project HBO22). Support from COST Action MP0806 is kindly
acknowledged.  The research leading to these results has received
funding from the European Research Council under the European
Community's Seventh Framework Program (FP7/2007-2013, Grant Agreement
no. 240579).

\

\bibliographystyle{jfm}
\bibliography{paperNotes}

\end{document}